\documentclass{jps-cp}
\usepackage{txfonts}
\usepackage{graphicx}
\newcommand{\qt}{q_{\perp}}

\title{Nuclear Modified Transverse Momentum Distributions and 3D Imaging in Nuclei}

\author{Mishary \textsc{Alrashed}$^{1}$, Daniele \textsc{Anderle}$^{2, 3, 1}$, Zhong-Bo \textsc{Kang}$^{1, 4, 5}$, John \textsc{Terry}$^{1, 4}$, and Hongxi \textsc{Xing}$^{2, 3}$}

\inst{$^{1}$Department of Physics and Astronomy, University of California, Los Angeles, California 90095, USA \\
$^{2}$Guangdong Provincial Key Laboratory of Nuclear Science, Institute of Quantum Matter,
South China Normal University, Guangzhou 510006, China \\
$^3$Guangdong-Hong Kong Joint Laboratory of Quantum Matter,
Southern Nuclear Science Computing Center, South China Normal University, Guangzhou 510006, China \\
$^4$Mani L. Bhaumik Institute for Theoretical Physics,
University of California, Los Angeles, California 90095, USA \\
$^5$Center for Frontiers in Nuclear Science, Stony Brook University, Stony Brook, New York 11794, USA}

\email{misharyalrashed@g.ucla.edu, dpa@m.scnu.edu.cn, zkang@ucla.edu, johndterry@physics.ucla.edu, hxing@m.scnu.edu.cn}

\recdate{March 14, 2022}

\abst{In this proceeding, we review our recent work in which we performed the first global analysis of nuclear modified Transverse Momentum Distribution Functions (TMDs). We demonstrate for the first time that the global set of TMD experimental data from HERMES, E866, E772, RHIC, ATLAS, and CMS, can be described using a simple model which accounts for the nuclear modifications for the TMDs as a non-perturbative correction. Using this model, we extract the nuclear modified TMDs for the first time.}

\kword{transverse momentum distributions (TMDs), nuclear medium, QCD global analysis, non-perturbative QCD}

\begin{document}
\maketitle

\section{Introduction} Transverse Momentum Dependent Parton Distribution Functions (TMD PDFs) encode three-dimensional information for the distribution of partons within hadrons, and provide vital information needed for imaging free nucleons. Due to the non-perturbative nature of QCD, global extractions from experimental data or lattice methods are required in order to obtain the full information of TMDs. Over the past decade tremendous advancements have been made in imaging hadron structure through intense experimental and theoretical studies in unpolarized and polarized TMDs, see for instance Refs.~\cite{EIC1,EIC2}. However, recently there has been additional interest in not only understanding TMDs for free nucleons, but also nuclear modified TMDs (nTMDs). In this proceeding, we review our recent global analysis for nTMDs using the world data from Semi-Inclusive DIS (SIDIS) and Drell-Yan processes with nuclei in Ref.~\cite{nTMDs}. We present detailed information of our methodology and results. This proceeding is organized as follows: in Sec.~\ref{sec:TMD} we provide the factorization and resummation formalism for SIDIS and Drell-Yan collisions with protons. In Sec.~\ref{sec:Proc}, we provide the details of our global fitting procedure. In Sec.~\ref{sec:Results} we summarize the results of our study. We conclude this proceeding in Sec.~\ref{sec:Conclusion}

\section{TMD Factorization Formalism}\label{sec:TMD}
The differential cross section for SIDIS, $e(l)+p(P)\rightarrow e(l')+h(P_h)+X$, in the TMD region for a proton target is given by the expression
\begin{align}
    \label{eq:fac-SIDIS}
    \frac{d\sigma}{dx_B\, dQ^2\, dz_h\, d^2P_{h\perp}} & = \sigma_0^{\rm DIS}\,H^{\rm DIS}(Q)\, \sum_q e_q^2 \int_0^\infty \frac{b\, db}{2\pi} J_0\left(b\, q_\perp\right) \,  D_{h/q}(z_h,b;Q)\,f_{q/p}(x_B,b;Q)\,.
\end{align}
In this expression, $b$ is the Fourier conjugate variable to the transverse momentum of the fragmenting quark, which is defined as $q_\perp = P_{h\perp}/z_h$. Furthermore, we can define the momentum of the virtual photon to be given by $q = l-l'$. The standard kinematic variables can be written in terms of this momentum as 
\begin{align}
    Q^2 = -q^2\,, \qquad
    x_B = \frac{Q^2}{2 P\cdot q}\,, \qquad
    z_h = \frac{P\cdot P_h}{P\cdot q}\,.
\end{align}
Furthermore in Eq.~\eqref{eq:fac-SIDIS}, $\sigma_0^{\mathrm{DIS}}$ and $H^{\rm DIS}$ are the usual Born cross section and the hard function for this process. The functions $f_{q/p}(x_B, b;Q)$ and $D_{h/q}(z_h, b;Q)$ denote the TMD PDF and the TMD Fragmentation Function (TMD FF), which provides three-dimensional information for a quark of flavor $q$ to form a hadron of species $h$. We note that in our notation both the TMD PDF and the TMD FF depend on the hard scale $Q$. In writing the TMDs in this form, we have set both the renormalization scale and the rapidity scale to be equal to $Q$. Using the usual Collins-Soper-Sterman formalism, these TMDs can be written as \cite{TMDs}
\begin{align}
\label{eq:TMDPDF}
    f_{q/p}(x_B,b;Q) = & \left[C_{q\leftarrow i}\otimes f_{i/p}\right] (x_B, \mu_{b_*})\, \exp\left\{-S_{\rm pert}(\mu_{b_*},Q) - S_{\rm NP}^f(x_B;Q_0,\mu_{b_*},Q)\right\},
    \\
    \label{eq:TMDFF}
    D_{h/q}(z_h,b;Q) = & \frac{1}{z_h^2}[\hat{C}_{i\leftarrow q}\otimes D_{h/i}] (z_h, \mu_{b_*}) \exp\left\{-S_{\rm pert}(\mu_{b_*},Q) - S_{\rm NP}^D(z_h;Q_0,\mu_{b_*},Q)\right\}\,.
\end{align}
In these expressions, $C_{q\leftarrow i}$ and $\hat{C}_{i\leftarrow q}$ are the Wilson coefficient functions for the TMD PDF and the TMD FF, which match the TMDs onto their collinear counterparts. In these expressions, $\otimes$ denotes the convolution operator, while $f_{i/p}(x, \mu_{b_*})$ and $D_{h/i}(z, \mu_{b_*})$ are the collinear PDF and collinear FF. The scale $\mu_{b_*} = 2e^{-\gamma_E}/b_*$ represents the natural scale for TMD evolution, while $b_*$ is the standard prescription. We note that since the convolutions are evaluated at the natural scale $\mu_{b_*}$, these convolution integrals require contributions from DGLAP evolution when computed for different $b$. Furthermore, in these expressions, $S_{\rm pert}$ is the perturbative Sudakov term associated with the TMD evolution. Finally, the $S_{\rm NP}^f$ and $S_{\rm NP}^D$ functions are the non-perturbative Sudakov factors associated with evolving the TMDs from the initial scale $Q_0$ to the hard scale $Q$. In the next section, we will provide our parameterization for these functions. We also note at this time that the goal of this study is to obtain these contributions for the nuclear TMDs.

The Drell-Yan differential cross section for $pp$ collisions, $p(P_1)+p(P_2)\rightarrow \gamma^*/Z(q)+X$, is given in the TMD region as
\begin{align}
    \label{eq:fac-DY}
    \frac{d\sigma}{dQ^2\, dy\, d^2\qt}  = \, \sigma_0^{\rm DY}\, & H^{\rm DY}(Q)\, \mathcal{P}\left(\eta, p_\perp^{\ell\ell}\right)\, \sum_q c_q(Q) \\
    & \times \int_0^\infty \frac{b\, db}{2\pi} J_0\left(b\, q_\perp\right) \, f_{\bar{q}/p}(x_1,b;Q)\, f_{q/p}(x_2,b;Q)\,.\nonumber
\end{align}
In this expression, $Q^2$, $y$, and $\qt$ denote the invariant mass, rapidity, and transverse momentum of the produced vector boson. Furthermore, $c_q(Q)$ denotes the quark coupling to the produced vector boson. The term $\mathcal{P}$ takes into account the kinematic cuts on the transverse momentum, $p_\perp^{\ell\ell}$, and the rapidity, $\eta$, of the final state lepton pair.

\section{Global Fitting Procedure}\label{sec:Proc}

In this section, we first discuss the available experimental data for the global analysis. We then discuss our parameterization for the non-perturbative physics. We note at this time, that we work at NLO+NNLL perturbative accuracy for all the perturbative physics.

Experimental measurement were performed at HERMES in Ref.~\cite{HERMES} for the multiplicity ratio $R^A_h = M^A_h/M^D_h$, where the multiplicity is defined as $M^A_h = 2\pi P_{h\perp} \frac{d\sigma^A}{dx\, dQ^2\, dz\, d^2P_{h\perp}}/ \frac{d\sigma^A}{dx dQ^2}$. The superscript of this expression denotes the species of the nuclear target and the subscript denotes the species of the produced final-state hadron. Experimental measurements were performed for both $\pi$ and $K$ production at HERMES. In our analysis however, we consider only the $\pi$ data and leave the analysis for the $K$ data for future work. We note that in order to calculate the multiplicities, one needs to compute the inclusive DIS cross section. To accomplish this, we use the APFEL library in Ref.~\cite{APFEL}. We also note at this time that we take the following cuts to select data within the TMD region: $P_{h \perp}^2 < 0.3\, \textrm{GeV}^2 $, and $z<0.7$. On the other hand, measurements of the Drell-Yan differential cross section ratio, $R_{\rm AB} = \frac{d\sigma^A}{dQ^2\, dy\, d\qt}/ \frac{d\sigma^B}{dQ^2\, dy\, d\qt}$, were performed by the E772 \cite{E772}, E866 \cite{E866}, and PHENIX \cite{RHIC} collaborations. Finally, experimental measurements of the $\qt$ distribution, $\frac{d\sigma^{\rm Pb}}{d\qt}$, were performed by both the ATLAS \cite{ATLAS} and CMS \cite{CMS} collaborations in $\rm{p\, Pb}$ collisions. To implement the fiducial cuts on these data, we use the Artemides library in Ref.~\cite{Art}. Finally, we note for Drell-Yan data, we will always take the kinematic cuts $\qt/Q<0.3$. In total, from the global set of data, we are left with 126 points.  

As we have seen in the previous paragraph, in order to describe experimental data, we must first parameterize the non-perturbative physics for protons. In this work, we follow the parameterization in Ref.~\cite{Params} to define the non-perturbative Sudakov to be
\begin{align}
    S_{\rm NP}^f(x_h;Q_0,\mu_{b_*},Q) & =  g_q\, b^2+\frac{g_2}{2}\ln\left(\frac{b}{b_*}\right)\ln\left(\frac{Q}{Q_0}\right)\,,
\\
    S_{\rm NP}^D(z_h;Q_0,\mu_{b_*},Q) & =  \frac{g_h}{z_h^2}\, b^2+\frac{g_2}{2}\ln\left(\frac{b}{b_*}\right)\ln\left(\frac{Q}{Q_0}\right)\,.
\end{align}
In Ref.~\cite{Params}, the parameter values were found from a global analysis of Semi-Inclusive DIS and Drell-Yan data to be $g_q = 0.106$ GeV$^2$, $g_h = 0.042$ GeV$^2$, $g_2 = 0.84$, and $Q_0 = \sqrt{2.4}$ GeV. Finally, in order to parameterize the collinear PDF, we use the CT14nlo parameterization in Ref.~\cite{CT14}. To parameterize the collinear FF, we use the DSS14 parameterization in Ref.~\cite{DSS}. 

To obtain the factorized cross section for interactions involving nuclei, we follow the same procedure that was used for both nuclear collinear PDFs and FFs in Refs.~\cite{EPPS,LIKEn}. Firstly, we assume that TMD factorization for nuclei is the same as that for free nucleons, except that one replaces the distributions by their nuclear modified versions. Secondly, we assume that the perturbative physics for TMDs inside the nucleus and the free nucleon is the same. Under this assumption, the Wilson coefficient functions and the perturbative Sudakov term are left unchanged. Furthermore, the DGLAP evolution kernels which enter into the convolution integrals are also left unchanged. While the perturbative contributions to the cross sections remain unchanged, we follow the work that was done for the collinear PDFs and FFs. Namely we treat the nuclear modifications to enter only into the non-perturbative parameterizations. To parameterize the nuclear modified collinear PDF and FF, we use the EPPS16 and LIKEn parameterization in Refs.~\cite{EPPS,LIKEn}. To parameterize the non-perturbative Sudakov, we take 
\begin{align}
\label{eq:nuclear-broad}
g_q^A(x, Q) =\, g_q + a_N\, L\,,
\quad
g_h^A(z, Q) =\, g_h + b_N\, L\,.
\end{align}
In this expression $L = A^{1/3}$, where $A$ is the atomic mass number, represents the transverse distance of the nucleus. Furthermore, $a_N$ and $b_N$ represent non-perturbative parameters to be tuned from experimental data involving nuclei. In order to obtain the numerical values of the parameters $a_N$ and $b_N$, we fit the experimental data using the Minuit package. The normalization factors $\cal{N}$ of the LHC data are accounted for in the definition of the $\chi^2$ according to the procedure of \cite{EPPS}.

\section{Results}\label{sec:Results}
\begin{figure*}[htb!]
    \centering
    \includegraphics[width = 1 \textwidth]{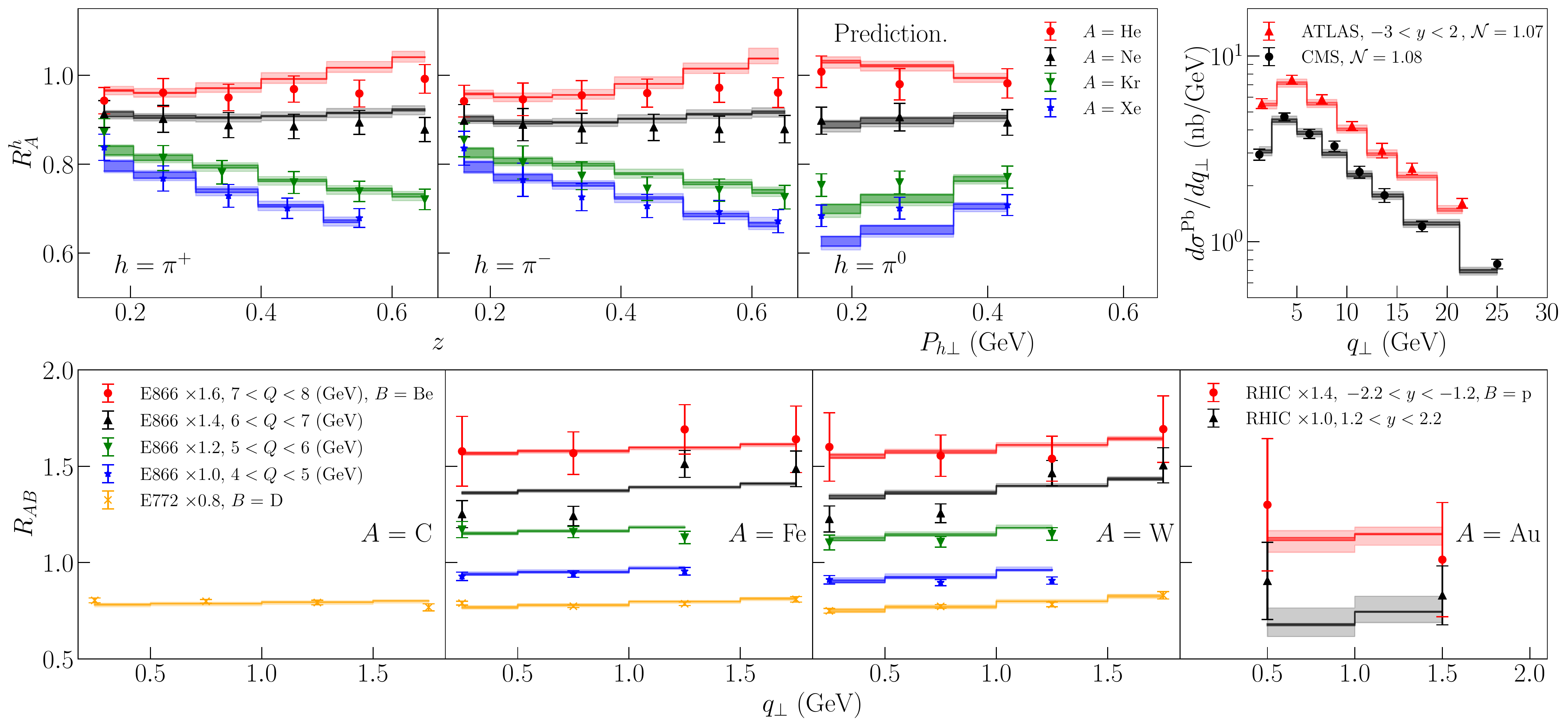}
    \caption{Our description of the considered experimental data. The dark band represents the uncertainty from our fit while the light band represents the uncertainty from the collinear distributions.}
    \label{fig:data-inc}
\end{figure*} 

The global analysis of these parameters results in a $\chi^2/d.o.f$ of $1.045$ where the parameter values are given by $a_N = 0.0171 \pm  0.003$ GeV$^2$ and $b_N = 0.0144 \pm 0.001$ GeV$^2$.  In Fig.~\ref{fig:data-inc}, we plot our description of the experimental data. The multiplicity ratio HERMES is plotted in the first three columns of the upper row. The $\qt$ distributions at the LHC are given in the upper right panel. In the bottom row, we plot our description of the $R_{AB}$ ratio from E772, E866, and RHIC experiments. In each subplot, we have provided the uncertainty from our fit as a dark band, and the uncertainty from the collinear distributions as a light band.

\begin{figure}[htb!]
    \centering\offinterlineskip
    \includegraphics[height = 0.2\textheight]{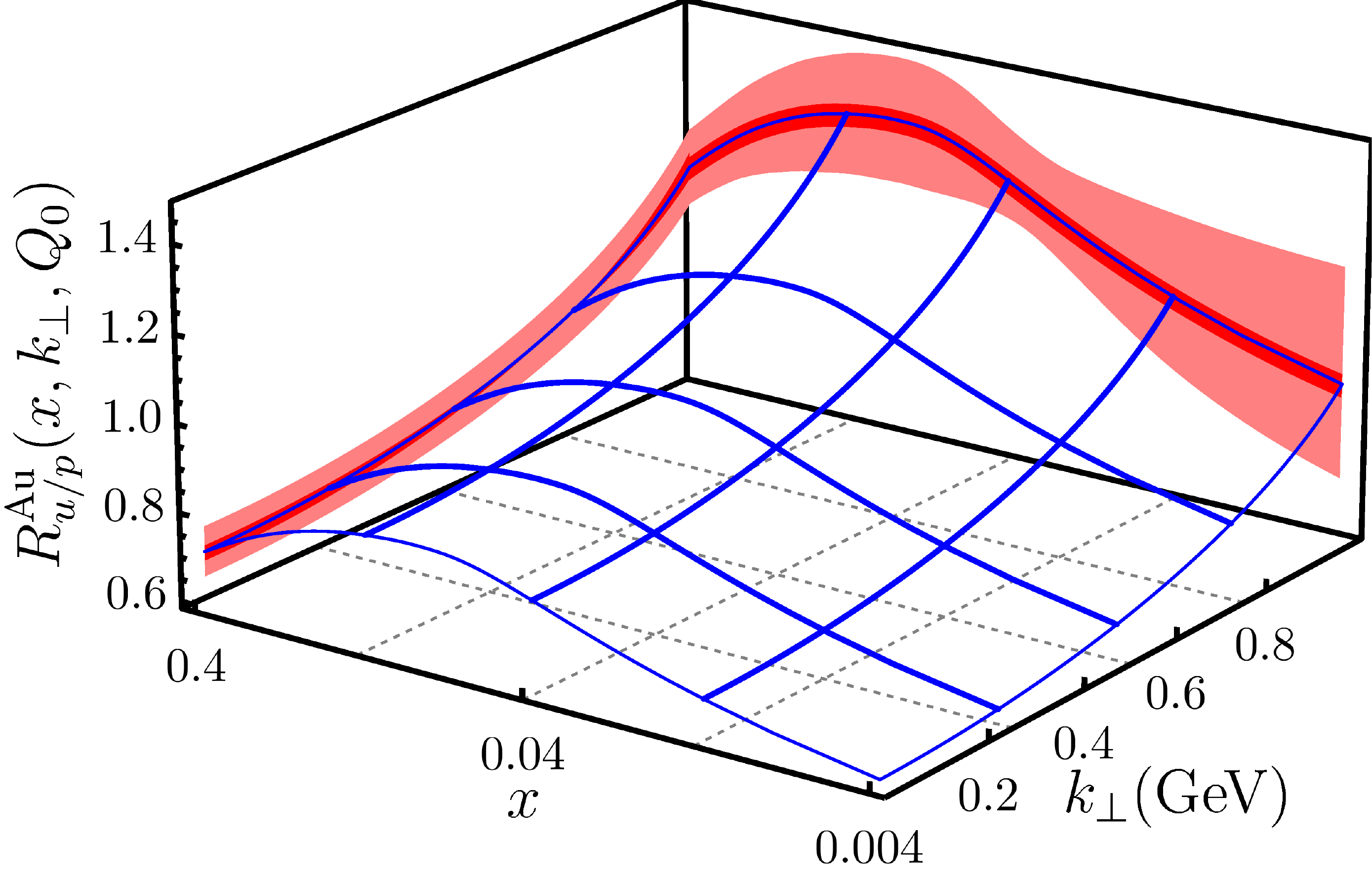}
    \includegraphics[height = 0.2\textheight]{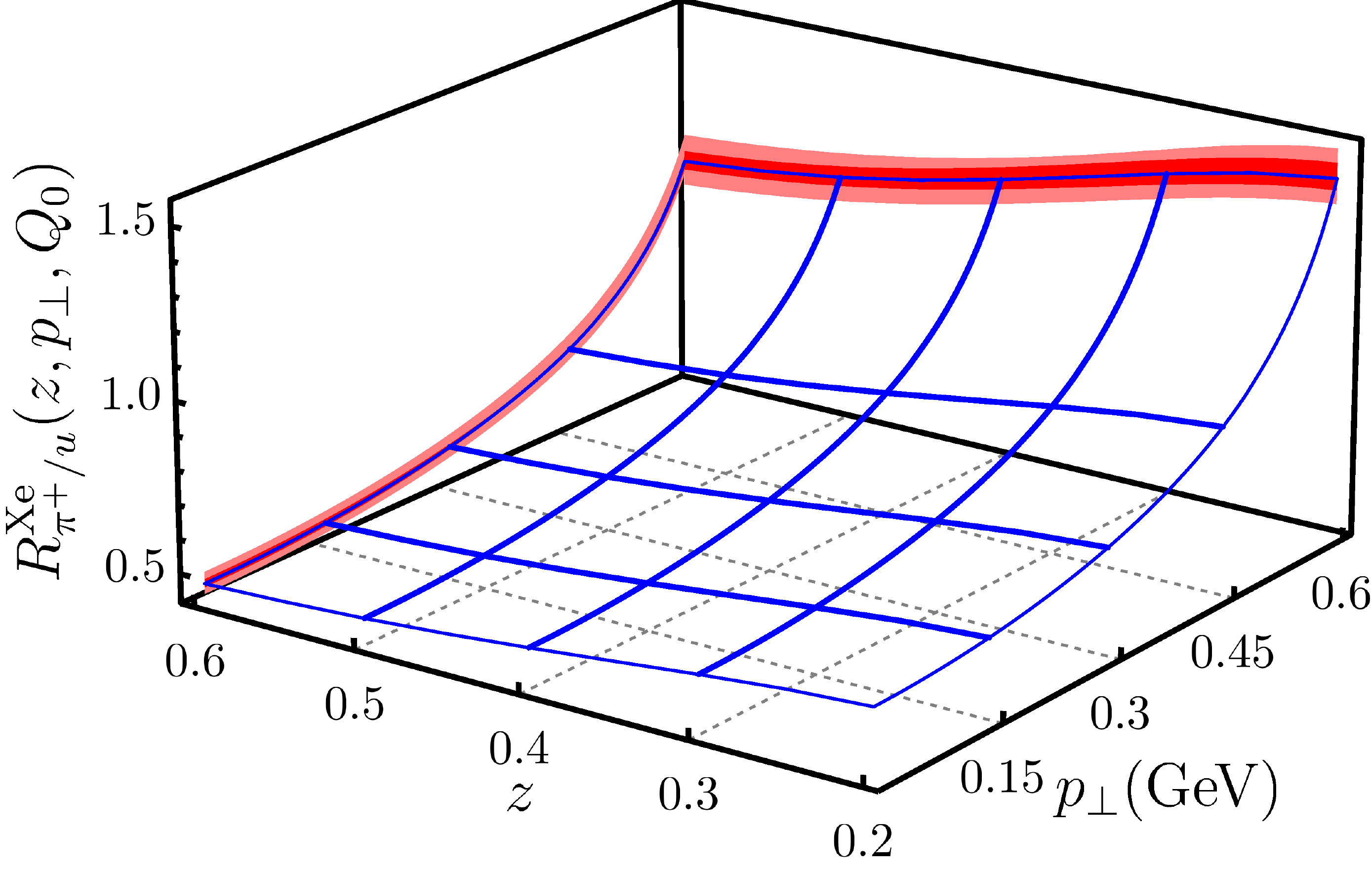}
  \caption{Left: The ratio of the $u$-quark TMD PDFs for a proton bounded in a $\rm Au$ nucleus to that in a free proton. Right: The ratio of the TMD FFs for a $u$-quark to fragment into a $\pi^+$ in the presence of a $\rm Xe$ nucleus.}
    \label{fig:nTMDs}
\end{figure}

In the left panel of Fig.~\ref{fig:nTMDs}, we plot the ratio of the $u$-quark TMD PDF of a bound proton in a gold nucleus and that in a free proton as a function of $x$ and $k_\perp$, the transverse momentum of the quark. In the right panel of this figure, we plot the ratio of the nTMD FF for $u\rightarrow \pi^+$ in a $\rm Xe$ nucleus and that in vacuum as a function of $z$ and $p_\perp$, the transverse momentum of the hadron with respect to the fragmenting quark. Analogous to the nTMD PDFs, we see that as $p_\perp$ grows, this ratio becomes larger, indicating that hadrons originating from fragmentation in the presence of a nuclear medium will tend to have a broader distribution of transverse momentum relative to vacuum TMD FFs. 

\section{Conclusion}\label{sec:Conclusion}

In this proceeding, we have reviewed our recent global analysis of nuclear modified TMDs. We have considered experimental data from HERMES, E772, E866, RHIC, and the LHC. We find that we can describe the global set of experimental data using a simple model which accounts for the nuclear modification as a non-perturbative effect.

\end{document}